\documentclass[]{spie}  
\usepackage[colorlinks,linkcolor=blue]{hyperref}
\usepackage{setspace}
 
\usepackage{amsmath,amsfonts,amssymb}
\usepackage{graphicx}
\usepackage{multirow}
\usepackage{subfigure}

\usepackage{booktabs}
\usepackage{bm}
\usepackage{hyperref}
\usepackage{float}

\hypersetup{hidelinks,
	colorlinks=true,
	allcolors=blue,
	pdfstartview=Fit,
	breaklinks=true}

\usepackage{setspace}

\title{Task-based Regularization in Penalized Least-Squares for Binary Signal Detection Tasks in Medical Image Denoising}

\author[a]{Wentao Chen}
\author[b]{Tianming Xu}
\author[c,d]{Weimin Zhou}

\affil[a]{University of Michigan-Shanghai Jiao Tong University Joint Institute, Shanghai Jiao Tong University, Shanghai, 200240, China}
\affil[b]{Global Institute of Future Technology, Shanghai Jiao Tong University, Shanghai, 200240, China}
\affil[c]{Wyant College of Optical Sciences, University of Arizona, Tucson, AZ 85721, USA}
\affil[d]{Department of Medical Imaging, University of Arizona, Tucson, AZ 85721, USA}

\authorinfo{Further author information: (Send correspondence to Weimin Zhou.)\\Weimin Zhou: E-mail: weiminzhou@arizona.edu}

\begin{document} 
	\maketitle
	
\begin{abstract}
Image denoising algorithms have been extensively investigated for medical imaging. To perform image denoising, penalized least-squares (PLS) problems can be designed and solved, in which the penalty term encodes prior knowledge of the object being imaged. Sparsity-promoting penalties, such as total variation (TV), have been a popular choice for regularizing image denoising problems. However, such hand-crafted penalties may not be able to preserve task-relevant information in measured image data and can lead to oversmoothed image appearances and patchy artifacts that degrade signal detectability. Supervised learning methods that employ convolutional neural networks (CNNs) have emerged as a popular approach to denoising medical images. However, studies have shown that CNNs trained with loss functions based on traditional image quality measures can lead to a loss of task-relevant information in images. Some previous works have investigated task-based loss functions that employ model observers for training the CNN denoising models. However, such training processes typically require a large number of noisy and ground-truth (noise-free or low-noise) image data pairs. In this work, we propose a task-based regularization strategy for use with PLS in medical image denoising. The proposed task-based regularization is associated with the likelihood of linear test statistics of noisy images for Gaussian noise models. The proposed method does not require ground-truth image data and solves an individual optimization problem for denoising each image. Computer-simulation studies are conducted that consider a multivariate-normally distributed (MVN) lumpy background and a binary texture background. It is demonstrated that the proposed regularization strategy can effectively improve signal detectability in denoised images.
\end{abstract}
	

\section{Purpose}
\label{sec:intro}  
Images produced by medical imaging systems are often subject to noise that are introduced in data acquisition process. 
Motivated by a common fact that many objects of interest can be described by a sparse representation, 
penalized least-squares (PLS) solutions associated with sparsity-promoting penalties can sometimes be employed for image noise reduction. Various sparsity-promoting penalties, such as total variation (TV) \cite{rudin1992nonlinear, tian2011low}, were explored to exploit \emph{a prior} information for regularizing the PLS problems. However, such hand-crafted penalties are often used to describe global information about the object and may not be aware of task-specific information related to clinically relevant tasks (e.g., tumor detection). 
For example, regularization of the PLS with TV penalty (PLS-TV) may produce images having small root-mean-square-error (RMSE) but introduce patchy artifacts that degrade signal detectability in images \cite{sidky2021signal}.

Supervised learning-based methods that employ deep neural neural networks (DNNs) have been extensively investigated to denoise medical images \cite{chen2017low, zhang2017beyond}. Such methods often train DNN denoisers by minimizing loss functions that measure an averaged distance between a group of noisy images and the corresponding ground-truth images (i.e., noise-free or low-noise images). It has been demonstrated that the trained DNN denoisers can achieve high denoising performance measured by traditional image quality metrics, such as RMSE and structural similarity index metric (SSIM). However, studies have shown that the DNNs trained with traditional supervised learning strategy can lead to degradation of task-based image quality measures associated with observer performance in signal detection tasks, and suggested the development of new training strategies that incorporate task-relevant information \cite{li2021assessing}. Recently, observer-based loss functions have been proposed to train DNNs for denoising medical images \cite{han2021low, li2022task}. However, such training-based methods require access to an ensemble of image data pairs of noisy and ground-truth images that may not be always available. 

In this work, we introduce a novel task-based regularization strategy for use with PLS denoising methods. Specifically, the proposed regularization term measures a distance between test statistics of the denoised images and the input noisy images. For Gaussian noise problems, this regularization corresponds to the log-likelihood of the test statistic of the noisy image when a linear observer is employed. 
Numerical studies are conducted that consider PLS-TV denoising problems involving a stochastic binary texture model and a multivariate-normally distributed lumpy background model. 
It is demonstrated that the noise-reduced images produced by the proposed task-based PLS-TV achieve significantly improved signal detectability, measured by both visual examination and receiver operating characteristic (ROC) analysis.

\section{Method}
\subsection{Binary signal detection tasks and model observers}
In this work, we consider a binary signal detection task in which an imaging system produces medical images that should be interpreted by an observer as satisfying a signal-absent hypothesis ($H_0$) or a signal-present hypothesis ($H_1$). The imaging processes under the hypotheses $H_0$ and $H_1$ can be described as \cite{barrett2013foundations}:
\begin{equation}
\begin{split}
    &H_{0}: \mathbf{g}=\mathbf{b}+\mathbf{n},\\
    &H_{1}: \mathbf{g}=\mathbf{b}+\mathbf{s}+\mathbf{n}.
    \end{split}
\end{equation}
Here, $\mathbf{g} \in \mathbb{R}^N$ denotes the measured noisy digital image, $\mathbf{b}\in\mathbb{R}^{N}$ and $\mathbf{s}\in\mathbb{R}^{N}$ denote the background image and signal image, respectively, and $\mathbf{n}\in\mathbb{R}^N$ is the measurement noise that degrades the image quality. 

To perform the binary signal detection task, a model observer can be employed that maps the measured digital image $\mathbf{g}$ to a test statistic, which is a real-valued scalar variable to be used in decision-making process in which it is compared to a predetermined threshold. Linear model observers, including the ideal linear observer (i.e., Hotelling observer) and anthropomorphic channelized Hotelling observers, have been widely used as an effective tool for objective assessment of image quality. Linear model observers compute test statistics as:
\begin{equation}
    t_{lin}(\mathbf{g})=\mathbf{w}^{T}\mathbf{g},
    \label{ts}
\end{equation}
where $\mathbf{w}\in \mathbb{R}^N$ is the observer template. In this work, we propose a regularization term that is guided by $t_{lin}(\mathbf{g})$ for use with PLS-TV in medical image denoising, which will be introduced in the following subsection.

\subsection{Task-based penalized least-squares for image denoising}
To address image denoising problems, a PLS optimization problem that involves a sparsity-promoting penalty can sometimes be solved:
\begin{equation}
    \hat{\mathbf{f}}=\arg\min_{\mathbf{f}}\left\{\|\mathbf{f}-\mathbf{g}\|_2^2+\lambda\cdot\Phi(\mathbf{f})\right\},
    \label{loss_traditional}
\end{equation}
where $\mathbf{f}$ is the sought-after clean image, $\|\mathbf{f}-\mathbf{g}\|_2^2$ is the data fidelity term, $\Phi(\mathbf{f})$ is the sparsity-promoting regularization or penalty that incorporates \emph{a prior} information about the expected solution, and $\lambda$ is the regularization parameter. It should be noted that the ground-truth image $\mathbf{f}\equiv \mathbf{b}$ under the $H_0$ hypothesis while $\mathbf{f}\equiv \mathbf{b + s}$ under the $H_1$ hypothesis. Popular sparsity-promoting penalties include $l_1$ norm of wavelet coefficients \cite{cai2012image} and total variation (TV) semi-norm \cite{osher2005iterative}. However, such handcrafted penalties can introduce oversmoothed and/or patchy appearance and may not be able to preserve task-relevant information in images for signal detection tasks (e.g., tumor detection). To address this problem, we propose a novel task-based regularization $\Psi_\mathbf{g}(\mathbf{f})$ to aid PLS optimization problem for preserving task-relevant information. The regularization $\Psi_\mathbf{g}(\mathbf{f})$ is designed based on the test statistic of a linear model observer:
\begin{equation}
\Psi_\mathbf{g}(\mathbf{f}) = \|t_{lin}(\mathbf{g})-t_{lin}(\mathbf{f})\|_2^2 = \|\mathbf{w}^T\mathbf{g}-\mathbf{w}^T\mathbf{f}\|_2^2.
\end{equation}
Here, the computation of observer template $\mathbf{w}$ is based on a training dataset of $\mathbf{g}$ and depends on specific linear observers selected (e.g., Hotelling observer, channelized Hotelling observer, and non-prewhitening matched filter). The final task-based PLS optimization problem can be described as:
\begin{equation}
    \hat{\mathbf{f}}=\arg\min_{\mathbf{f}}\{\alpha\cdot\|\mathbf{f}-\mathbf{g}\|_{2}^2+\beta\cdot \Phi(\mathbf{f})+\gamma\cdot \Psi_\mathbf{g}(\mathbf{f})\},
    \label{loss}
\end{equation}
where $\alpha$, $\beta$, and $\gamma$ are tunable parameters.

It should be noted that the proposed regularization $\Psi_\mathbf{g}(\mathbf{f})$ is closely related to the likelihood function of observer's test statistics. When the noise $\mathbf{n}\sim\mathcal{N}\left(\mathbf{0},\mathbf{K}_{\mathbf{n}}\right)$ follows a Gaussian distribution with zero mean and covariance matrix of $\mathbf{K}_{\mathbf{n}}$, the probability density function of $t_{lin}(\mathbf{g})$ conditioned on $\mathbf{f}$ is Gaussian:  
\begin{equation}
    \Pr(t_{lin}(\mathbf{g}) = t \mid \mathbf{f}) = \Pr(\mathbf{w}^T\mathbf{g} = t \mid \mathbf{f}) = \Pr(\mathbf{w}^T(\mathbf{f} + \mathbf{n}) = t \mid \mathbf{f}) = \Pr(\mathbf{w}^T\mathbf{n} = t - \mathbf{w}^T\mathbf{f}).
\end{equation}
It can be shown that $\mathbf{w}^T\mathbf{n}$ is a Gaussian random variable with mean of zero and variance of $\mathbf{w}^T\mathbf{K}_{\mathbf{n}}\mathbf{w}$ and subsequently, $\Pr(t_{lin}(\mathbf{g}) = t \mid \mathbf{f})$ is a Gaussian density function with the mean of $\mathbf{w}^T\mathbf{f}$ and variance of $\mathbf{w}^T\mathbf{K}_{\mathbf{n}}\mathbf{w}$. Consequently, the task-based regularization is proportional to the negative log-likelihood $\log(\Pr(t_{lin}(\mathbf{g}) = t \mid \mathbf{f}) )$. In this preliminary study, we choose the TV semi-norm as the sparsity-promoting penalty $\Phi(\mathbf{f})$ and employ the Hotelling template $\mathbf{w}_{HO}$ to form the task-based regularization term $\Psi_\mathbf{g}(\mathbf{f})$.

\section{Numerical studies and results}
\subsection{Dataset and simulation setup}
Computer-simulation studies that consider two binary signal detection tasks were conducted in which the following stochastic object models were employed.

\begin{figure}[!ht]
\centering
\subfigure{
    \begin{minipage}[t]{0.14\linewidth}
 	\centering
 	\includegraphics[width=\textwidth]{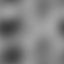}\\
 	\end{minipage}
}
\subfigure{
    \begin{minipage}[t]{0.14\linewidth}
 	\centering
 	\includegraphics[width=\textwidth]{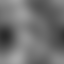}\\
 	\end{minipage}
}
\subfigure{
    \begin{minipage}[t]{0.14\linewidth}
 	\centering
 	\includegraphics[width=\textwidth]{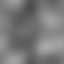}\\
 	\end{minipage}
}
\subfigure{
    \begin{minipage}[t]{0.14\linewidth}
 	\centering
 	\includegraphics[width=\textwidth]{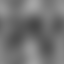}\\
 	\end{minipage}
}
\subfigure{
    \begin{minipage}[t]{0.14\linewidth}
 	\centering
 	\includegraphics[width=\textwidth]{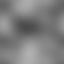}\\
 	\end{minipage}
}
\subfigure{
    \begin{minipage}[t]{0.14\linewidth}
 	\centering
 	\includegraphics[width=\textwidth]{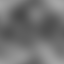}\\
 	\end{minipage}
}
\centering
\caption{Examples of the generated MVNLumpy background images.}
\label{mvnvis1}
\end{figure}

\noindent\textbf{MVNLumpy object model:} We utilized a stochastic multivariate-normally distributed lumpy model (MVNLumpy), also known as the type-2 lumpy model \cite{rolland1990factors}, to generate background images with dimension 64 $\times$ 64. Fig. \ref{mvnvis1} shows examples of the generated MVNLumpy background images. A Gaussian signal with a standard deviation of 5 and magnitude of 0.02 was added to the background to simulate signal-present images. A Gaussian noise with a mean of zero and a standard deviation of 0.01 was added to simulate the noisy image data.

\begin{figure}[H]
\centering
\subfigure{
    \begin{minipage}[t]{0.14\linewidth}
 	\centering
 	\includegraphics[width=\textwidth]{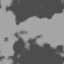}\\
 	\end{minipage}
}
\subfigure{
    \begin{minipage}[t]{0.14\linewidth}
 	\centering
 	\includegraphics[width=\textwidth]{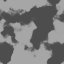}\\
 	\end{minipage}
}
\subfigure{
    \begin{minipage}[t]{0.14\linewidth}
 	\centering
 	\includegraphics[width=\textwidth]{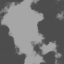}\\
 	\end{minipage}
}
\subfigure{
    \begin{minipage}[t]{0.14\linewidth}
 	\centering
 	\includegraphics[width=\textwidth]{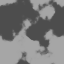}\\
 	\end{minipage}
}
\subfigure{
    \begin{minipage}[t]{0.14\linewidth}
 	\centering
 	\includegraphics[width=\textwidth]{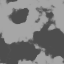}\\
 	\end{minipage}
}
\subfigure{
    \begin{minipage}[t]{0.14\linewidth}
 	\centering
 	\includegraphics[width=\textwidth]{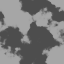}\\
 	\end{minipage}
}
\centering
\caption{Examples of the generated binary texture background images.}
\label{binmvnvis1}
\end{figure}

\noindent\textbf{Binary texture model:} In this case, we employed a binary texture model \cite{abbey2008ideal} that was designed to mimic breast CT textures. To generate a background $\mathbf{b}$, a thresholding operation was applied element-wisely to a Gaussian random field. In our work, we utilized a Gaussian $1/f$ noise and sigmoid threshold function to generate binary texture background of size 64 $\times$ 64. Examples of the generated binary texture images are shown in Fig. \ref{binmvnvis1}. A circular disk signal with radius 2 and magnitude 0.07 was added to the background at the center of images for simulating signal-present objects. To generate noisy measurement data, we added a Gaussian noise with mean of zero and standard deviation of 0.1.

\noindent\textbf{Simulation setup:}
The PLS and task-based PLS optimization problems were solved by use of Adam optimizer \cite{kingma2014adam} in PyTorch framework \cite{paszke2019pytorch}. The Adam optimizer with a learning rate of $1\times 10^{-4}$ was deployed to minimize the loss function. We set the number of iterations for the optimization process to 10,000. To form the task-based regularization in the PLS problem, we calculated the Hotelling template $\mathbf{w}_{HO}$ using 190,000 noisy signal-absent and 190,000 noisy signal-present images. Additionally, for binary texture model, we trained a denoising convolutional neural network (DnCNN) \cite{zhang2017beyond} on 10,000 noisy/clean signal-absent image pairs and 10,000 noisy/clean signal-present image pairs that was used in comparison studies. During the training of the DnCNN, the learning rate was set to $1\times 10^{-3}$, and the mean squared error loss function was minimized with Adam optimizer.

\subsection{Results}
We applied $\mathbf{w}_{HO}$ to a test dataset that comprised 2,000 signal-absent images and 2,000 signal-present images to evaluate signal detection performance, which is quantified by the area under the receiver operating characteristic (ROC) curve (AUC) value.

\begin{figure}[!ht]
\centering
\subfigure[\label{roc_mvn}]{
    \begin{minipage}[t]{0.5\linewidth}
 	\centering
 	\includegraphics[width=\textwidth]{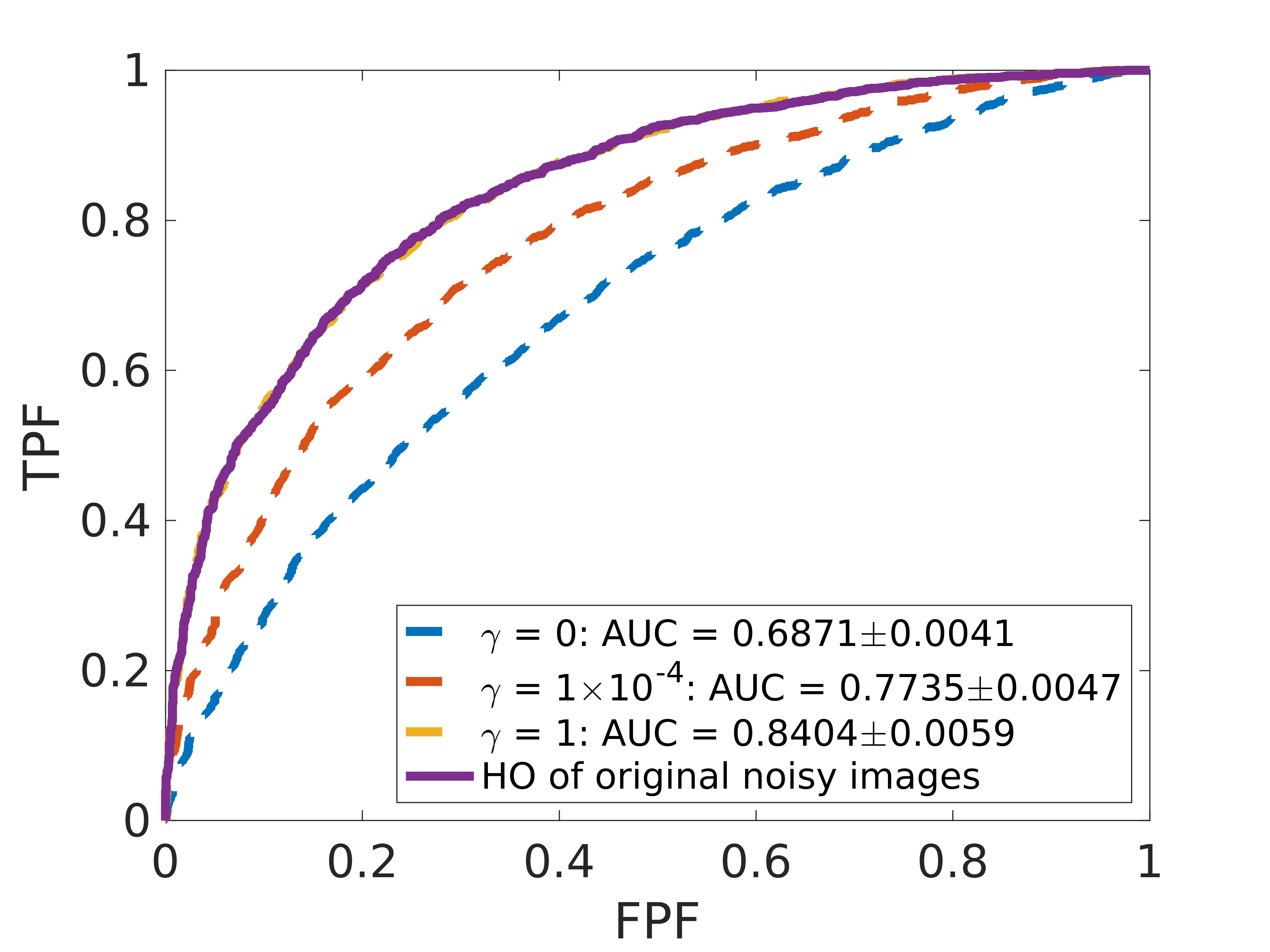}\\
 	\end{minipage}
}
\subfigure[\label{auc_binmvn}]{
    \begin{minipage}[t]{0.45\linewidth}
 	\centering
 	\includegraphics[width=\textwidth]{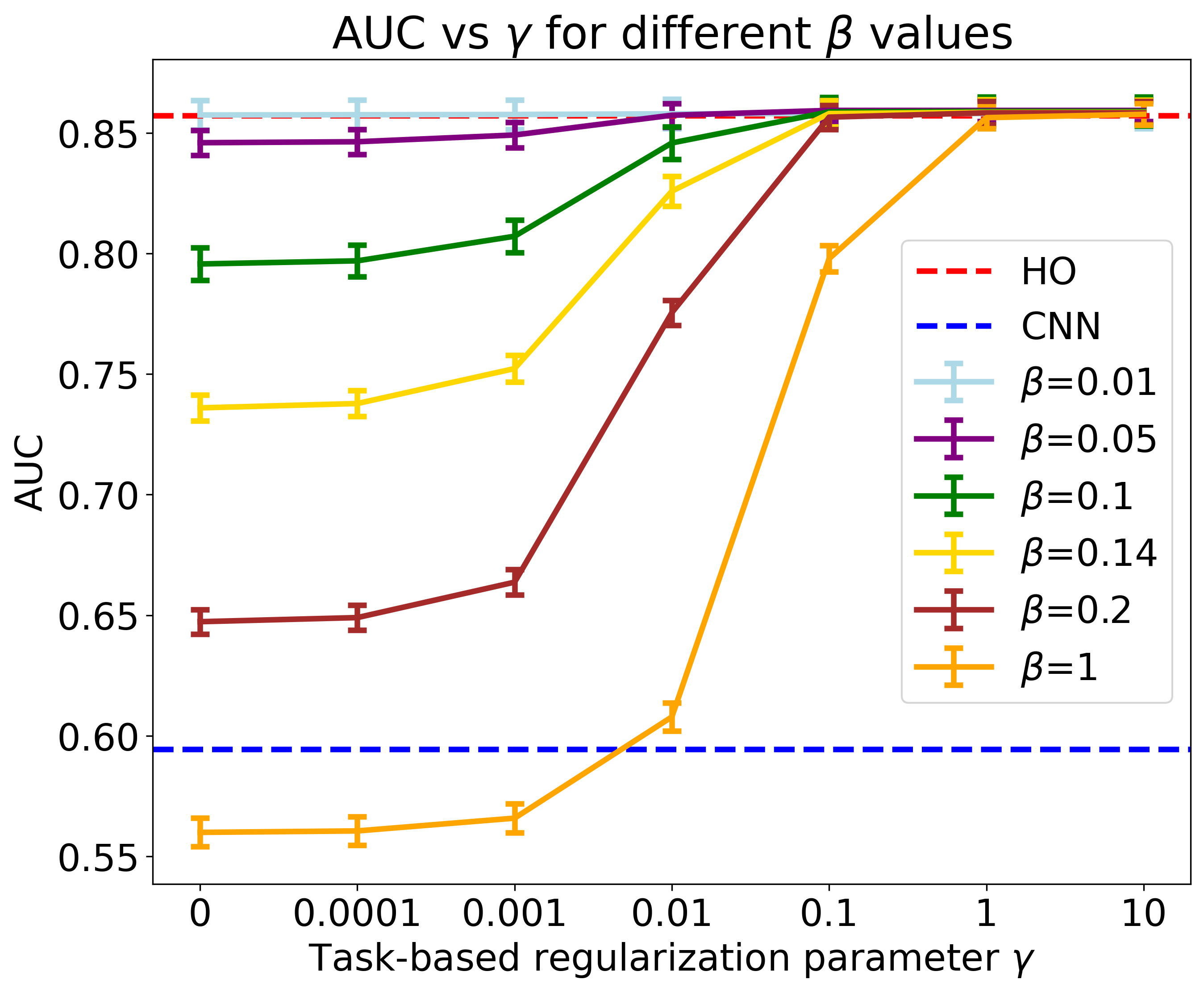}\\
 	\end{minipage}
}
\centering
\caption{(a) The ROC curves produced by denoised images using different $\gamma$ values and original noisy images on the MVNLumpy object model. (b) Evaluation results on the binary texture model: the AUC values corresponding to denoised images with different total variation regularization parameters $\beta$ and different task-based regularization parameters $\gamma$, and the AUC values corresponding to the original noisy data and CNN denoised images are plotted.}
\label{binmvnvis}
\end{figure}

For MVNLumpy object model, we set the parameters $\alpha=1$, $\beta=0.05$, and $\gamma$ to various values between 0 and 1. The ROC curves corresponding to different $\gamma$ values are shown in Fig. \ref{roc_mvn}. As the $\gamma$ gets larger, the binary signal detection performance corresponding to the denoised images improved and became closer to the detection performance corresponding to the original noisy images.

In the study with binary texture model, we applied our method with the parameters $\alpha=1$ and different $\beta$ values ranging from 0.01 to 1. Each $\beta$ value combined with various $\gamma$ values from 0 to 10. The performance of the Hotelling observer acting on the original noisy images and the denoised images produced by use of different parameters are compared in Fig. \ref{auc_binmvn}. The AUC value was decreased when the TV regularization parameter $\beta$ was increased. Additionally, as expected, the AUC value was increased when the task-based regularization parameter $\gamma$ was increased.

\begin{figure}[H]
\centering
\subfigure[Ground-truth\label{binmvnbs}]{
    \begin{minipage}[t]{0.18\linewidth}
 	\centering
 	\includegraphics[width=\textwidth]{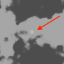}\\
 	\end{minipage}
}
\subfigure[Noisy\label{binmvnbsn}]{
    \begin{minipage}[t]{0.18\linewidth}
 	\centering
 	\includegraphics[width=\textwidth]{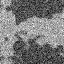}\\
 	\end{minipage}
}
\subfigure[CNN denoised\label{binmvncnn}]{
    \begin{minipage}[t]{0.18\linewidth}
 	\centering
 	\includegraphics[width=\textwidth]{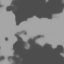}\\
 	\end{minipage}
}
\subfigure[$\gamma=0$\label{binmvn0}]{
    \begin{minipage}[t]{0.18\linewidth}
 	\centering
 	\includegraphics[width=\textwidth]{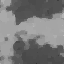}\\
 	\end{minipage}
}
\subfigure[$\gamma=0.0001$\label{binmvn00001}]{
    \begin{minipage}[t]{0.18\linewidth}
 	\centering
 	\includegraphics[width=\textwidth]{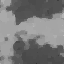}\\
 	\end{minipage}
}
\subfigure[$\gamma=0.001$\label{binmvn0001}]{
    \begin{minipage}[t]{0.18\linewidth}
 	\centering
 	\includegraphics[width=\textwidth]{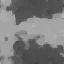}\\
 	\end{minipage}
}
\subfigure[$\gamma=0.01$\label{binmvn001}]{
    \begin{minipage}[t]{0.18\linewidth}
 	\centering
 	\includegraphics[width=\textwidth]{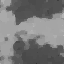}\\
 	\end{minipage}
}
\subfigure[$\gamma=0.1$\label{binmvn01}]{
    \begin{minipage}[t]{0.18\linewidth}
 	\centering
 	\includegraphics[width=\textwidth]{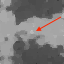}\\
 	\end{minipage}
}
\subfigure[$\gamma=1$\label{binmvn1}]{
    \begin{minipage}[t]{0.18\linewidth}
 	\centering
 	\includegraphics[width=\textwidth]{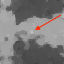}\\
 	\end{minipage}
}
\subfigure[$\gamma=10$\label{binmvn10}]{
    \begin{minipage}[t]{0.18\linewidth}
 	\centering
 	\includegraphics[width=\textwidth]{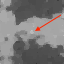}\\
 	\end{minipage}
}
\centering
\caption{(a) ground-truth signal-present image; (b) noisy signal-present image; (c) denoised image using DnCNN; (d) - (j) denoised images using proposed method under different $\gamma$ values with $\alpha=1$ and $\beta=0.14$.}
\label{binmvnresult}
\end{figure}

A visual examination study was conducted with $\alpha=1$ and $\beta=0.14$. Fig. \ref{binmvnbs} and Fig. \ref{binmvnbsn} display the original ground-truth image and the corresponding noisy image, respectively. Fig. \ref{binmvncnn} shows the denoised images produced by the DnCNN. It can be observed that although the denoised image produced by the DnCNN is clean and preserves most textures, it loses task-relevant information about the signal to be detected, which substantially impacts the signal detection performance. This issue also appeared in the PLS method that solely employs TV penalty, as illustrated in Fig. \ref{binmvn0}. The denoised images produced by our proposed task-based PLS method are shown in Fig. \ref{binmvn00001} to Fig. \ref{binmvn10}. It can be observed that the denoised images produced by the task-based PLS with appropriately selected $\gamma$ (e.g., 0.1, 1, and 10) possess a more accurate representation of the signal, which leads to a significant improvement of signal detection performance, as illustrated in Fig. \ref{auc_binmvn}. 

\begin{figure}[H]
\centering
\subfigure[$\gamma=0.001$]{
    \begin{minipage}[t]{0.18\linewidth}
 	\centering
 	\includegraphics[width=\textwidth]{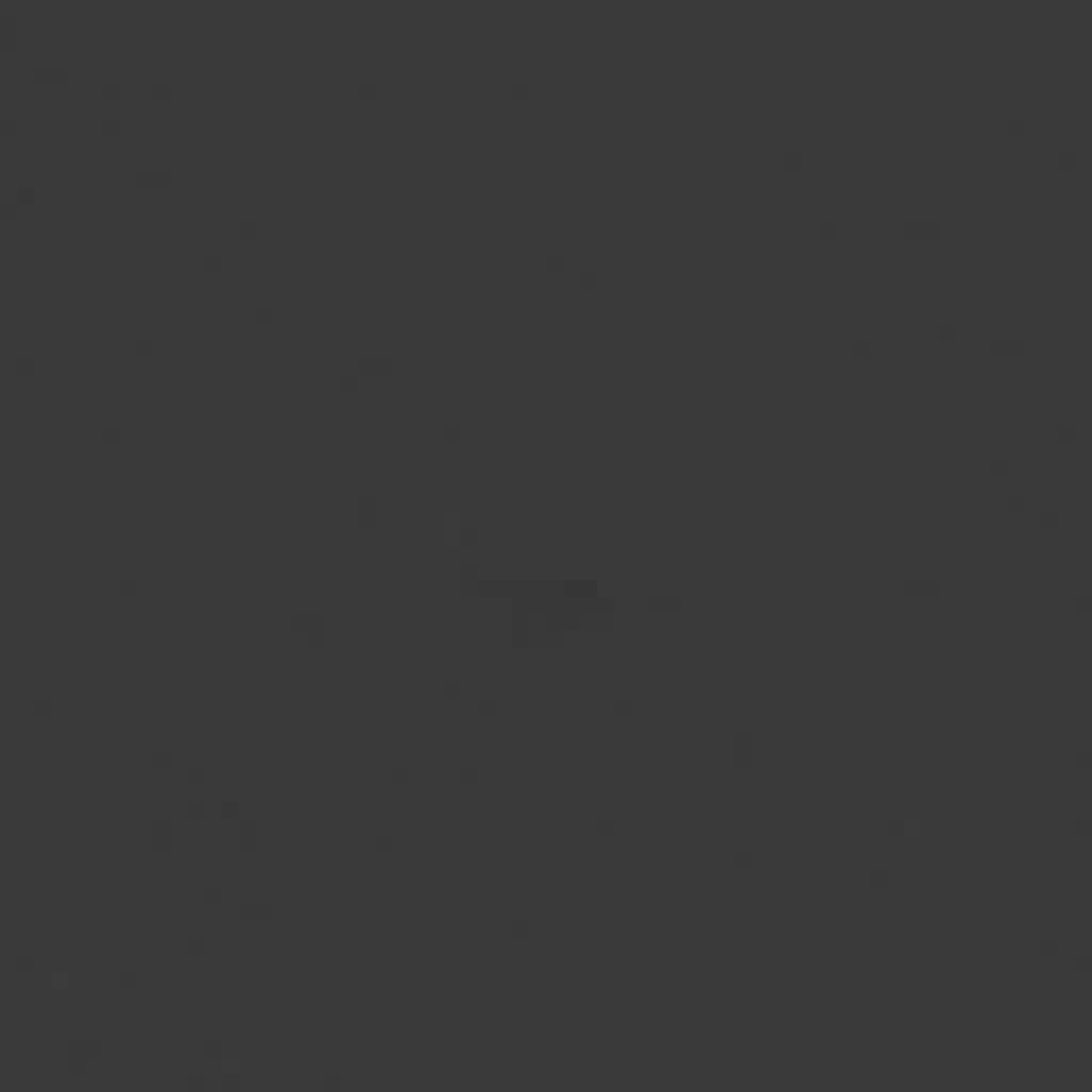}\\
 	\end{minipage}
}
\subfigure[$\gamma=0.01$]{
    \begin{minipage}[t]{0.18\linewidth}
 	\centering
 	\includegraphics[width=\textwidth]{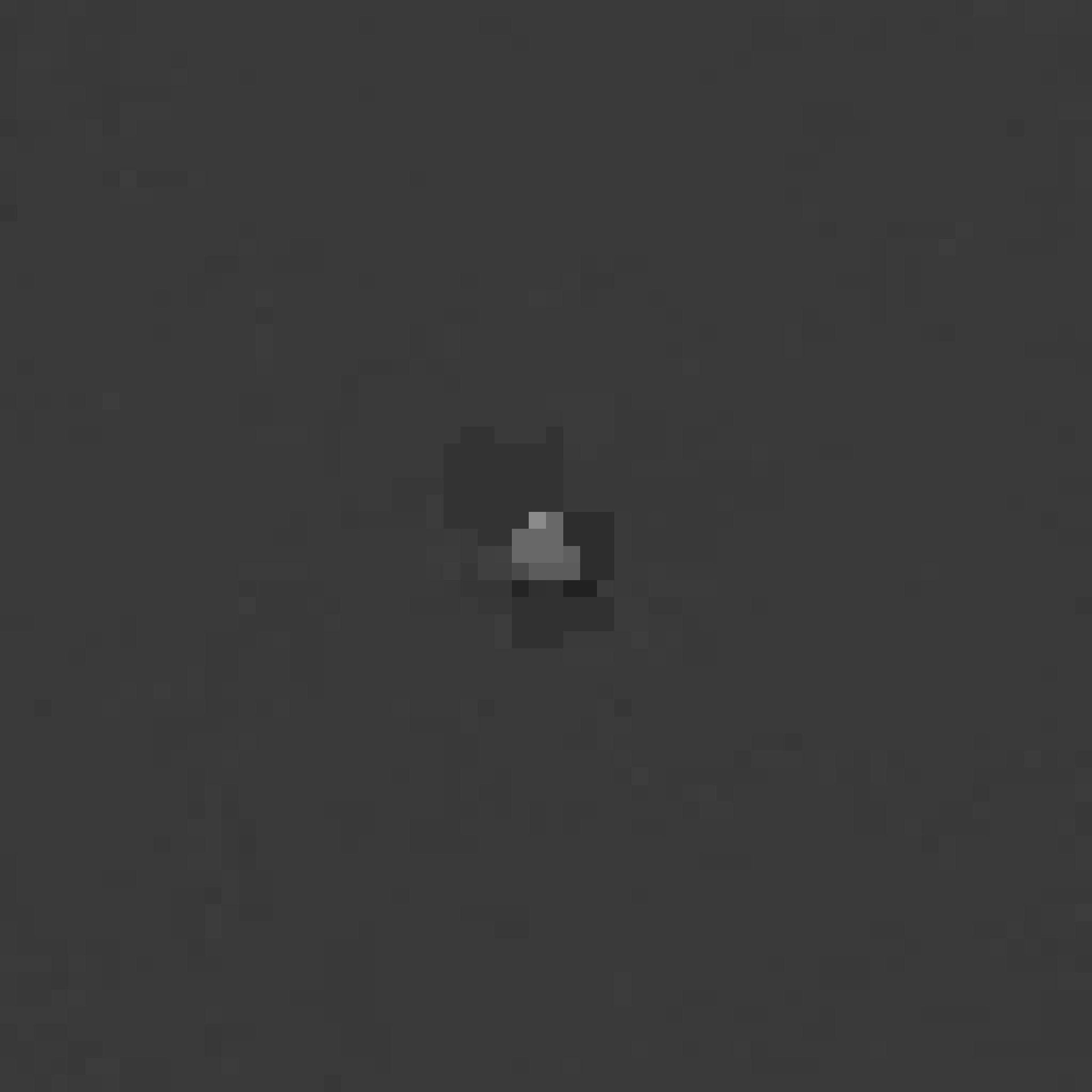}\\
 	\end{minipage}
}
\subfigure[$\gamma=0.1$]{
    \begin{minipage}[t]{0.18\linewidth}
 	\centering
 	\includegraphics[width=\textwidth]{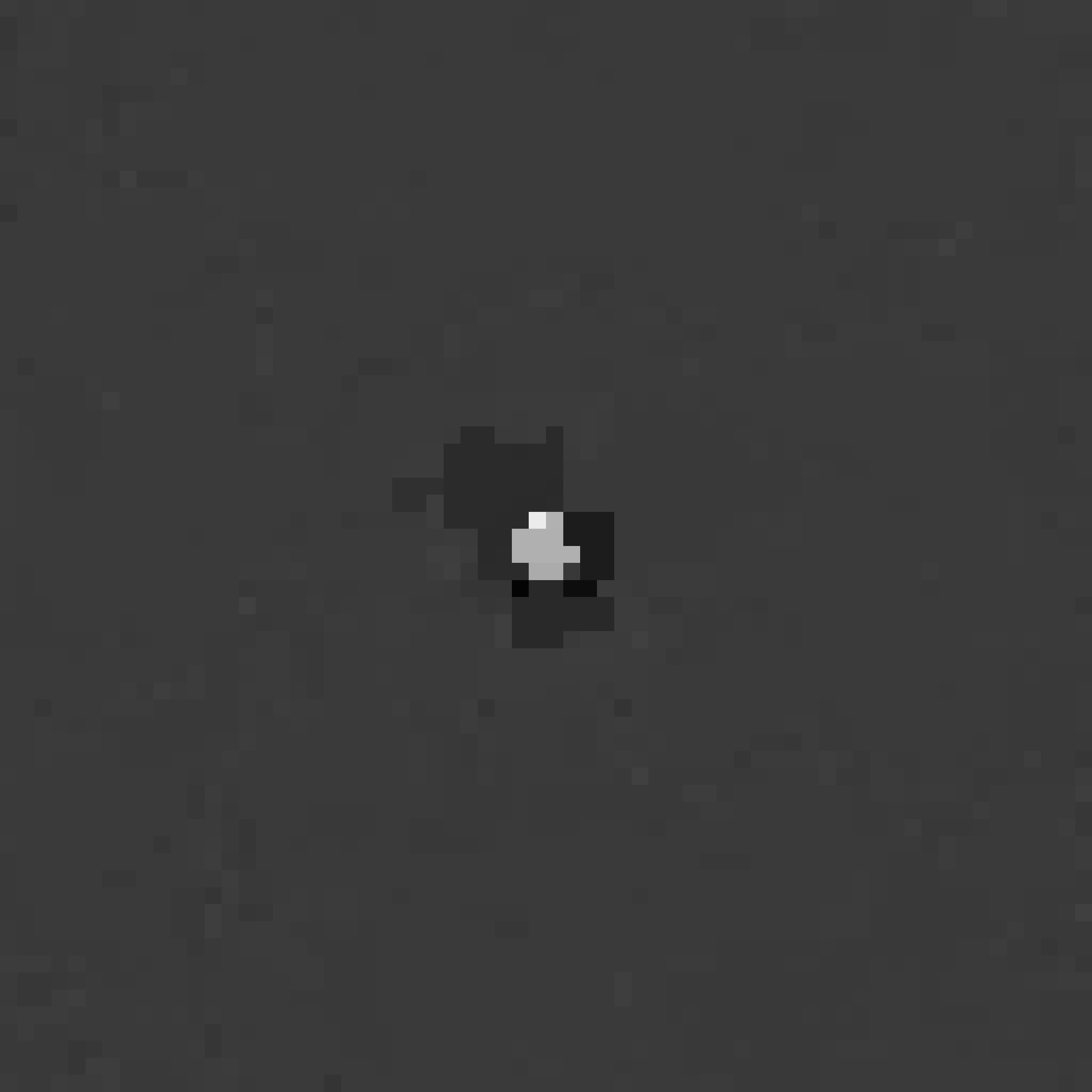}\\
 	\end{minipage}
}
\subfigure[$\gamma=1$]{
    \begin{minipage}[t]{0.18\linewidth}
 	\centering
 	\includegraphics[width=\textwidth]{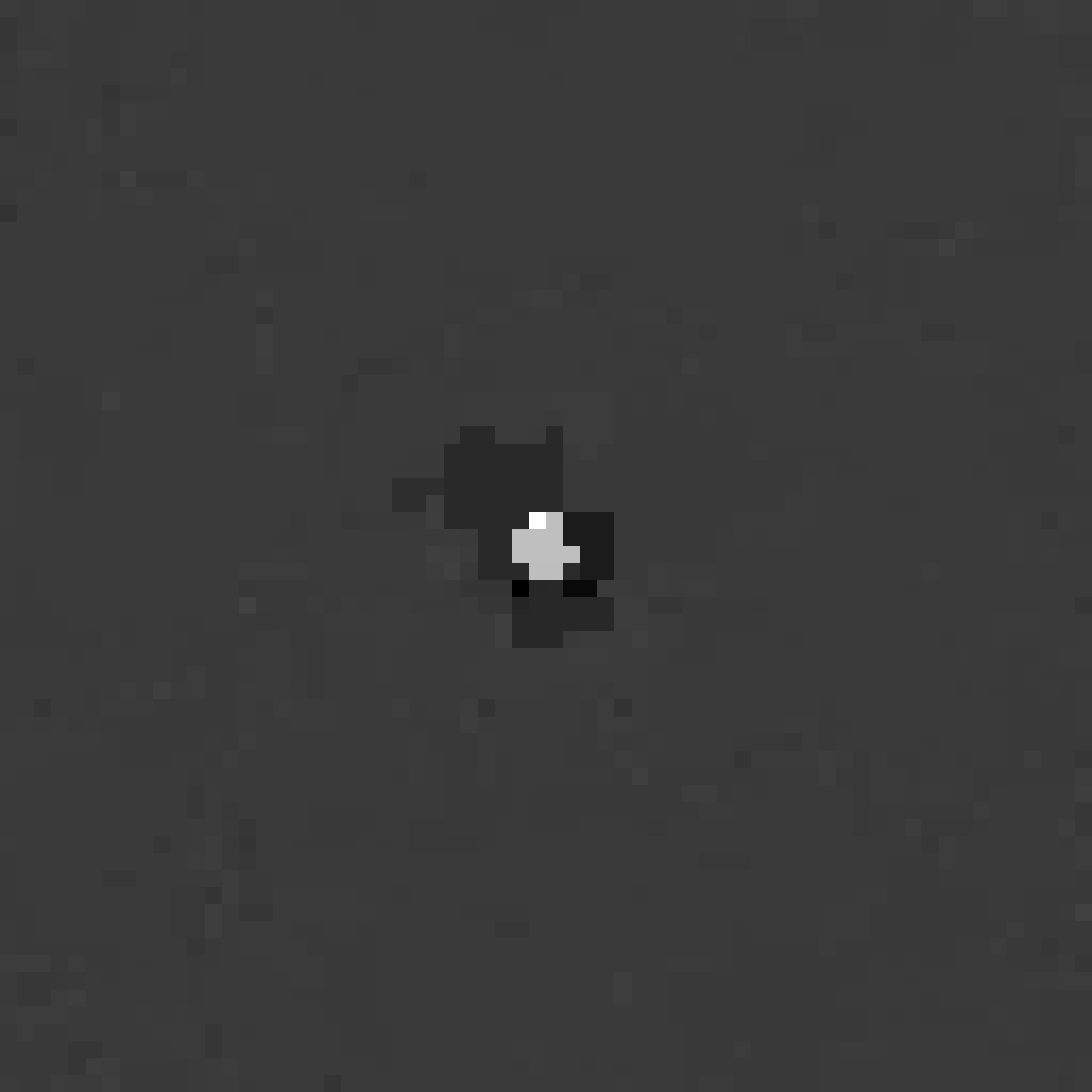}\\
 	\end{minipage}
}
\subfigure[$\gamma=10$]{
    \begin{minipage}[t]{0.18\linewidth}
 	\centering
 	\includegraphics[width=\textwidth]{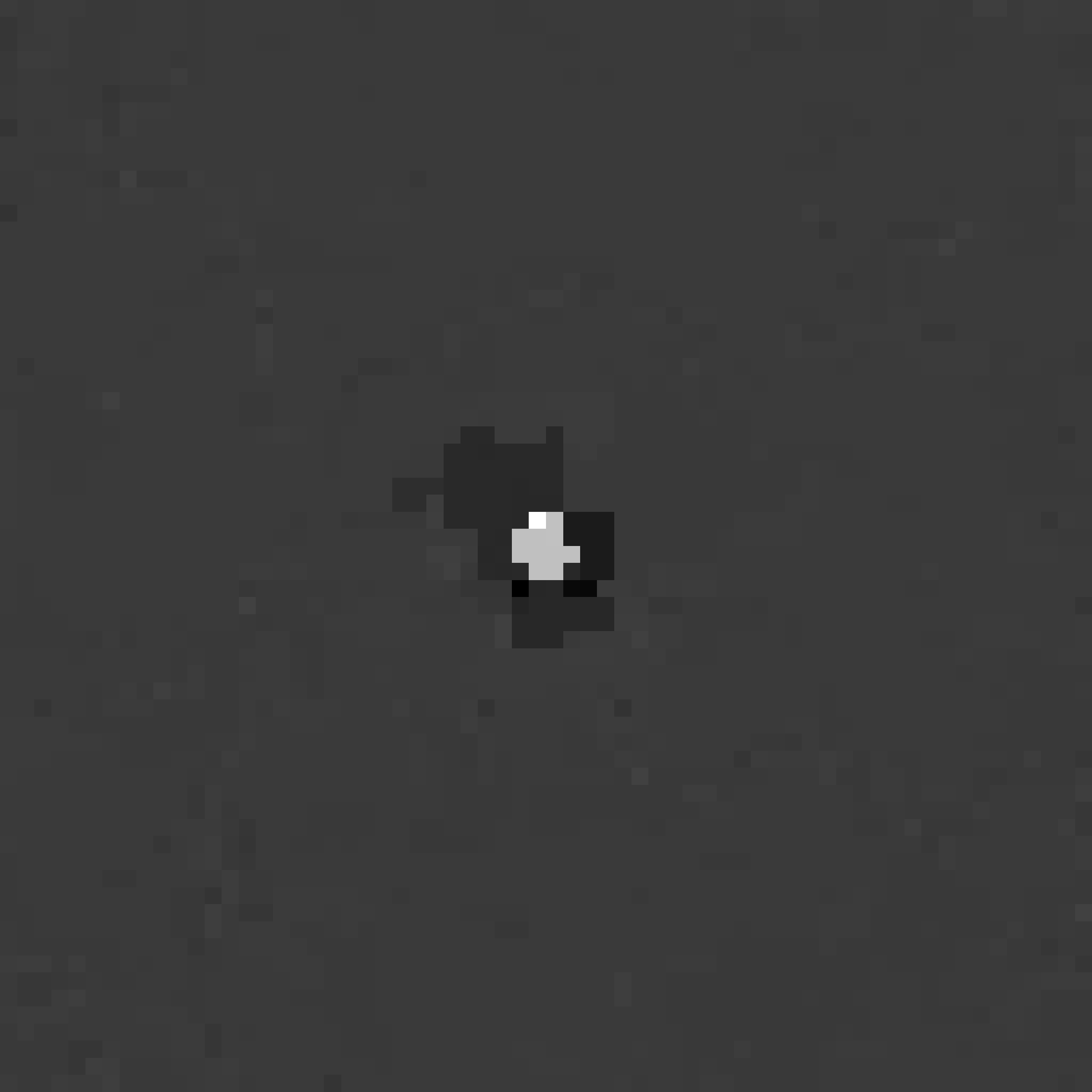}\\
 	\end{minipage}
}
\centering
\caption{Difference maps between denoised images produced by the task-based PLS-TV method ($\gamma>0$) and the traditional PLS-TV method ($\gamma=0$).}
\label{diff}
\end{figure}

To better understand the impact of the task-based regularization parameter $\gamma$ on preserving task-relevant information, difference maps between images associated with various positive $\gamma$ (task-based PLS-TV) and $\gamma = 0$ (original PLS-TV) under $\alpha=1$ and $\beta=0.14$ were calculated and are shown in Fig. \ref{diff}. As expected, the results demonstrate that when $\gamma$ increases, the task-relevant information can be better preserved in images and the signal to be detected can be easier perceived.

\section{Conclusions}
In this study, a task-based regularization strategy was proposed for use with penalized least-squares method for medical image denoising. The proposed regularization term corresponds to the likelihood of a linear test statistic when the image noise is Gaussian. Numerical studies that considered a MVNLumpy object model and a binary texture model were systematically conducted. The denoised images were evaluated via both visual examination study and task-based image quality analysis. It has been demonstrated that our proposed method can be highly effective in preserving task-specific information when performing medical image denoising for signal detection tasks. 
This preliminary study only considered using linear observer to form the task-based regularization term. In future works, we will investigate the use of deep learning-based non-linear ideal observers \cite{zhou2019approximating, zhou2020approximating, zhou2023Ideal} to regularize inverse problems for performing clinically relevant tasks in medical imaging.

\bibliography{report} 
\bibliographystyle{spiebib} 
	
\end{document}